\font\twelvebf=cmbx12
\font\ninerm=cmr9
\nopagenumbers
\magnification =\magstep 1
\overfullrule=0pt
\baselineskip=18pt
\line{\hfil }
\line{\hfil October 1997}
\vskip .8in
\centerline{\twelvebf  Comments on the 
Problem of a Covariant Formulation} 
\centerline{\twelvebf of Matrix Theory} 
\vskip .3in
\centerline{\ninerm H.AWATA and D.MINIC}
\vskip .1in
\centerline{Enrico Fermi Institute}
\centerline{The University of Chicago}
\centerline{Chicago, IL 60637}
\centerline{and}
\centerline{Physics Department}
\centerline{Pennsylvania State University}
\centerline{University Park, PA 16802}
\vskip .1in
\centerline {awata@yukawa.uchicago.edu, minic@hbar.phys.psu.edu}

\vskip 1in
\baselineskip=16pt
\centerline{\bf Abstract}
\vskip .1in

A possible avenue towards 
the covariant formulation of the bosonic Matrix Theory 
 is explored. The approach is guided
by the known covariant description of the bosonic membrane. We point
out various problems with this particular 
covariantization scheme, stemming from the central 
question of
how to enlarge the original $U(N)$ symmetry of Matrix Theory 
 while preserving all of its essential
features in the infinite momentum frame.

\vfill\eject

\footline={\hss\tenrm\folio\hss}

\magnification =\magstep 1
\overfullrule=0pt
\baselineskip=22pt
\pageno=1

Matrix theory, unlike its classical
counterpart - membrane theory [1], is currently formulated only 
in a background dependent
way [2]. This article addresses the question of whether a  
covariant formulation of the bosonic
part of Matrix theory is possible which 
 emulates the 
well-known covariant description of the bosonic membrane.
(Although we do not discuss
supersymmetry in this paper, 
our analysis is motivated in part by the fact
 that a covariant formulation of the full
supersymmetric Matrix theory would circumvent the puzzling issue
of longitudinal boost invariance, and point out
a way towards a covariant description of
 the Matrix theory five-brane.)

The basic idea of this note is 
to take the existing covariant action for a    
 $d+1$ dimensional bosonic 
membrane [1] (with, let's say, $d=10$), which possesses the 
full three-dimensional 
world-volume diffeomorphism invariance, and which 
 in the light-cone gauge 
leads to a Hamiltonian that is 
invariant under the residual area preserving
diffeomorphisms of the transverse membrane, and examine
its matrix analog. The light-cone membrane Hamiltonian
coincides with the bosonic Matrix Theory 
Hamiltonian [2], upon application of 
the Goldstone-Hoppe prescription 
[3]. This prescription can be interpreted as
a regularization of the
residual area preserving diffeomorphisms in terms of 
the $U(N)$ ($N = \infty$) rotations of matrices [3], the 
transverse spatial coordinates being mapped into $N \times N$ 
hermitian matrices, which describe the transverse coordinates
of $N$ $D0$-branes (the $U(N)$ symmetry governs the statistics
of a system of $N$ $D0$-branes [2],[4]).     

The following natural question arises of whether it is possible to 
 implement an analogous   
 regularization
procedure  
in the original
 covariant description of the bosonic
membrane. Then, in case one is fortunate
enough to start 
with a regularized covariant theory with sufficiently 
 large symmetry (some appropriate matrix analog of 
the three-dimensional diffeomorphisms) one could study the
light-cone gauge and determine whether 
 the light-cone Hamiltonian coincides with the 
bosonic Matrix theory
Hamiltonian.

The note is organized in three parts: To start with, we review some
well known facts about the covariant formulation of
the bosonic membrane following the classic reference [1].
We then present a possible route towards covariant formulation of Matrix
theory guided by the analogy with the covariant membrane 
dynamics. Finally, we discuss various problems with this approach, 
all of which are rooted in the crucial question of
how to extend the original $U(N)$ symmetry of
Matrix theory 
(or how to regularize the full three-dimensional 
diffeomorphisms of the covariant membrane formulation)  
 and be able to recover the infinite momentum
frame description upon gauge fixing.
We also compare our results with Smolin's recent analysis 
of 
the question of the covariant quantization
of membrane dynamics [6].

Let us then start from the familiar bosonic membrane action [1]
$$
S = - \int d^{3} \xi \sqrt{-det g_{ij}}, \eqno(1)
$$
where the induced world-volume metric $g_{ij}$ is 
$$
g_{ij} = \partial_{i} x_{\mu} \partial_{j} x_{\nu} \eta^{\mu \nu}, \eqno(2)
$$
and where $x_{\mu} (\mu= 0,1,...,d) $ 
denote the target space coordinates of a $d+1$ dimensional
bosonic membrane. $x_{\mu}$'s are functions of
the three
 world-volume coordinates $\xi^{i}, i=0,1,2$.

The world-volume reparametrization invariance 
$\delta x_{\mu} = \epsilon^{i} \partial_{i} x_{\mu}$
allows one to go
to light-cone gauge and rewrite the theory in terms of 
physically relevant transverse variables. The light-cone
coordinates are defined as 
$$
x_{\pm} = {1 \over \sqrt{2}} ( x_d \pm x_0), \eqno(3a)
$$
and   
the light-cone gauge     
$$
\partial_{i} x_{+} = \delta_{i0}. \eqno(3b)
$$   
The world volume coordinates $\xi^{i}$ split as
$$
(\xi^{0}, \xi^{s}) \rightarrow (t, \xi^{s}). \eqno(3c)  
$$

The Lagrangian density reads as follows
(we adopt the notation found in 
[1])     
$$
{\cal{L}} = - \sqrt{ g \Delta }, \eqno(4)
$$
where $g \equiv det g_{rs}$ ($r,s=1,2$) is the 
determinant of the induced two dimensional
metric, and $g$ and
 $\Delta$ are determined by     
$$\eqalign{
g &= {1 \over 2} \{x_{a}, x_{b}\}^{2} \cr
\Delta &= - g_{00} + u^{r} u_{r} \equiv - |D_{0} x_{\mu}|^{2}. 
} \eqno(5)
$$
Here $D_{0} \equiv \partial_{0} - u^{r} \partial_{r}$ 
and $u^{r} = g^{rs} u_{s}$
($u_{r}\equiv g_{0r}$
plays the role analogous to that of the shift function
in the Hamiltonian approach to general relativity),
 $a= 1,...,d-1$ stands for the index of the
transverse space, and $\{,\}$ denotes
the usual Poisson brackets with respect to
$\xi_{s}$   
$$
\{ x_{a},x_{b} \} \equiv \partial_{\xi^{1}} x_{a} \partial_{\xi^{2}}x_{b}
- \partial_{\xi^{2}} x_{a} \partial_{\xi^{1}} x_{b}.   
$$ 
 
 Furthermore, in the light-cone gauge (3b)
$$\eqalign{
g_{00} &= 2 {\partial}_{0}{x}_{-} + ({\partial}_{0}{x}_{a})^{2} \cr
u_{r}  &= \partial_{r} x_{-} + 
{\partial}_{0}{x}_{a} {\partial}_{r}{x}_{a}.  
} \eqno(6)
$$
The conjugate momenta
are easily calculated   
$$\eqalign{
p_{a} &= \sqrt{\Delta^{-1} g} 
({\partial}_{0}{x}_{a} -  u^{r} {\partial}_{r} x_{a}) =
  \sqrt{\Delta^{-1} g} D_{0} x_{a} \cr
p_{+} &= \sqrt{\Delta^{-1} g} .  
} \eqno(7)
$$

Then the light-cone Hamiltonian density  
$ {\cal{H}} = p_{a} \partial_{0} x_{a} + p_{+} \partial_{0} x_{-} - {\cal{L}}$
reads simply as follows
$$
{\cal{H}}
 = {1 \over 2 p_{+}} ({p_{a}}^{2} + {1 \over 2} \{x_{a}, x_{b}\}^{2}).   
\eqno(8)
$$
{}From this form of the light cone Hamiltonian it is seen that the 
original three-volume diffeomorphisms $\delta x_{\mu} = \epsilon^{i}
\partial_{i} x_{\mu}$ reduce to the area preserving
diffeomorphisms described by 
$\delta x_{a} = \epsilon^{rs} \partial_{r} w 
\partial_{s} x_{a} = \{ w , x_{a} \}$. The longitudinal
coordinate $x_{-}$ does not appear explicitly in (8) and is determined 
from the primary constraint $p_{a} \partial_{r} x_{a} + p_{+}
\partial_{r} x_{-} \sim 0$ and the requirement that the
longitudinal momentum is time-independent  
$\partial_{0} p_{+} =0$, or essentially,  
the gauge condition $u^{r} =0$ [1] .

At this point the light-cone Hamiltonian is regularized by applying
the Goldstone-Hoppe map between representation theories of the
algebra of the area preserving diffeomorphisms and 
the $N = \infty$ limit
of Lie algebras [3].
The Goldstone-Hoppe prescription instructs us to perform
 the following translation of the transverse spatial
coordinates   
$$
\{x_{a}, x_{b}\}  \rightarrow [X_{a}, X_{b}], \eqno(9a)
$$
where now $X_{a}$ denote large $N \times N$ hermitian
matrices (the area preserving diffeomorphisms are mapped into
$U(N)$ rotations of matrices 
for the simplest case of a spherical membrane).
Also
$$
\int d \xi^{1} d \xi^{2}  \rightarrow Tr.  \eqno(9b) 
$$
The resulting regularized Hamiltonian density is precisely
the bosonic part of the Matrix theory Hamiltonian [2].   
(The bosonic membrane thus represents a classical configuration
of the bosonic part of Matrix theory.)

After this short review let us return to the original covariant
bosonic membrane action (1) and try to apply the Goldstone-Hoppe
regularization at the very covariant level.
(The covariant formulation of the bosonic part of Matrix 
theory should thus contain the covariant bosonic membrane as a natural
classical configuration.)  
First, we notice that
the three by three determinant in (1) can
be expanded in such a way so that  
there appears a term of the form 
$$
{1 \over 2} (\partial_{t} x_{\mu})^{2} \{x_{\rho}, x_{\nu}\}^{2}, 
\eqno(10a) 
$$
and two other terms which can be collected together to read 
($\xi^{1} \equiv \sigma$, $\xi^{2} \equiv \tau$)
$$
- (\partial_{\sigma} x_{\nu} (\partial_{t} x^{\mu} \partial_{\tau} 
x_{\mu}) - 
\partial_{\tau} x_{\nu} (\partial_{t} x^{\mu} \partial_{\sigma} 
x_{\mu}))^{2}.
\eqno(10b)  
$$
In other words, the original Nambu-Goto type action  
for the bosonic membrane (1) can be written  
in the so-called Barbour-Bertotti form [5], [6] 
$$
S = - \int d^3 \xi \sqrt{-{1 \over 2} {\dot{x}}^{\mu}
{\dot{x}}^{\nu} (\eta_{\mu \nu} g - 2 \{x_{\mu},x_{\beta}\}
\{x_{\nu},x^{\beta}\})}. \eqno(10c)
$$
Here ${\dot{x}}_{\mu} \equiv \partial_{t} x_{\mu}$.
This action naturally incorporates time reparametrization invariance
on the world-volume of the membrane.

The application of the original Goldstone-Hoppe dictionary 
to (10c) then leads (at least naively) to the following regularized
action 
$$
S_{M} = - Tr \int dt 
\sqrt{-{1 \over 2} ({\dot{X}}_{\mu})^{2} [X_{\mu}, X_{\nu}]^{2}
+ ({\dot{X}}_{\mu} [X_{\mu}, X_{\nu}])^{2} }, \eqno(11)
$$    
where the Poisson brackets with respect to
$\xi^{s}$ get replaced by the commutators of
time-dependent matrices and
the integral over $\xi^{s}$ by the usual 
matrix trace $Tr$. $X_{\mu}(t) (\mu=0,...,d)$
represent now large $N \times N$ hermitian matrices;
$t$ plays the role of a "world-line" parameter. Obviously, to
properly define this expression one has to examine the question of 
ordering. One way to do this is to symmetrize all matrix products. Another
question pertains to the definition of
$\sqrt{M}$, where $M$ is an $N \times N$ matrix. We  
use one of the formal expressions, for example, 
$\sqrt{M} \equiv \exp({1 \over 2} \log{M})$ or $\sqrt{M} \equiv
(1 - (1 - M))^{1/2} = 1 - {1 \over 2}(1 - M) +... $, so that
$M^{n} M^{m} = M^{n + m}$ and $(u M u^{-1})^{n} = u M^{n} u^{-1}$, for
some unitary $N \times N$ matrix $u$.

What are the symmetries of (11)?
One immediately sees that apart from "world-line" $t$-reparametrization
invariance, (11) is endowed with the following symmetry
$$
X_{\mu} \rightarrow u X_{\mu} u^{-1} + f_{\mu}, \eqno(12)
$$
where $u$ is a unitary $t$-independent matrix (describing
unitary time-independent rotations), and 
$f_{\mu} \propto {\bf 1}$ (${\bf 1}$ is the unit $N \times N$ 
matrix) are constant matrices (describing
constant shifts). (Notice the similarity of (12) to a global,
time-independent, Poincare invariance.)
 It can be seen that 
(12) together with $t$-reparametrization symmetry is not
enough to try to go to light-cone gauge i.e. set
$X_{+} \propto {\bf 1}$ ;
here $X_{\pm} \equiv {1 \over \sqrt{2}} (X_{d} \pm X_{0})$. 

However, there exists another natural  
expression for $S_{M}$ which leads to completely equivalent physical results
in the continuum membrane limit, 
namely 
$$
S_{M} = - \int dt Tr \sqrt{-{1\over 2}
(D_{0}X_{\mu})^{2} [X_{\nu}, X_{\rho}]^{2}}, \eqno(13)
$$
where $D_{0} \equiv \partial_{0} - [\omega,.]$. Eq. (13) follows
from eqs. (1) - (6) and the Goldstone-Hoppe dictionary. $\omega$ is
the matrix analog of the shift function $u^{r}$ in (5).
 (Observe that both (11) and
(13) are natural on dimensional grounds.)
However, (13) 
is invariant 
under
$$
X_{\mu}(t) \rightarrow u(t) X_{\mu}(t) u^{-1}(t) + f_{\mu}, \eqno(14) 
$$
in addition to "world-line" 
$t$-reparametrization invariance $t \rightarrow \phi(t)$
(note that $u$ is now a time-dependent matrix; $f_{\mu} \propto {\bf 1}$ as
before).  
The first term in (14) describes time-dependent $U(N)$ rotations
and the second - translations. Also
$$
\omega \rightarrow u(t) \omega u^{-1}(t) + (\partial_{0} u) u^{-1}. \eqno(14a)
$$
 
Given (14) we can immediately diagonalize $X_{+}$
$$
X_{+} \sim diag(\phi_{1}(t),...,\phi_{N}(t)). \eqno(15a)
$$
Now comes the crucial point. One could expect that in the large N limit
all of $\phi_{i}(t) \rightarrow \phi(t)$. Not being able to
prove this statement, we take it as a crucial assumption and
proceed with the computation of the light-cone Lagrangian and
Hamiltonian! 
If we assume that $\phi_{i}(t) \rightarrow \phi(t)$,
by using $t$-reparametrization
invariance we can set $\phi(t) \rightarrow t$ (therefore
recovering the global time of Matrix Theory in the infinite
momentum frame), so that finally the
gauge condition reads       
$$
X_{+} = (X_{+}(0) + t) {\bf 1}. \eqno(15b)
$$
This equation would
define the matrix
version of the light-cone gauge (3b).

We can further fix the gauge (by utilizing (14a) and
letting $\omega = 0$, the matrix analog of $u^{r} = 0$ gauge [1]) thus  
ending up with  
$$
S_{M} = - \int dt Tr \sqrt{-{1\over2}(\partial_{0}X_{\mu})^{2} [X_{\nu}, 
X_{\rho}]^{2}}. 
\eqno(16)
$$
(Here $(\partial_{0} X_{\mu})^{2} = 
2{\dot{X}}_{-} + ({\dot{X}}_{a})^{2}$ and
$[X_{\mu}, X_{\nu}]^{2} = [X_{a},X_{b}]^{2}$.)  
This action is seen to be invariant under the residual transformations    
$$
X_{a}(t) \rightarrow \lambda X_{a}(t) \lambda^{-1} + f_{a}, \eqno(17) 
$$
where $\lambda$ is again a constant unitary matrix and
$f_{a} \propto {\bf 1}$.

To summarize: Starting from a covariant description of M-theory 
of a Nambu-Goto type (13) in
terms of $d+1$ time-dependent matrices, which is invariant under 
time-dependent $U(N)$ rotations
of matrices and constant translations
(14) as well as "world-line" $t$-reparametrizations, and by naively 
diagonalizing one of the eleven matrices, in order
to define the matrix analog of 
the "light-cone" gauge (15a,b) (which appears to be
 possible only in  the large $N$ 
limit
), we end up with 
(16). (In essence, we propose to regularize the volume
preserving part of the full three-dimensional
diffeomorphisms describing the covariant membrane dynamics by $t$-dependent
$U(N)$ rotations and "world-line" $t$-reparametrizations in the 
$N=\infty$ limit. We argue that the light-cone gauge fixing is
possible only in the large $N$ limit. Upon gauge fixing we
find that the light-cone
Lagrangian is invariant under the residual
symmetry (17).)

At this point we can (at least
formally) rewrite the Lagrangian density  
${\cal{L}}_{M}$ given by (16) as 
$$
{\cal{L}}_{M} \equiv  -\sqrt{\Delta } \sqrt{g}. \eqno(18)
$$
Here we adopt a particular ordering prescription for the Lagrangian, conjugate
momenta and Hamiltonian, in order to make contact with the Matrix theory
Hamiltonian description. 
The matrices $g$ and $\Delta$ are defined as follows
$$\eqalign{
g &= {1 \over 2} [X_{\mu}, X_{\nu}]^{2} = {1 \over 2} [X_{a},X_{b}]^{2} \cr
\Delta &= -(D_{0} X_{\mu})^{2} \rightarrow - (\partial_{0} X_{\mu})^{2}
= - 2 {\dot{X}}_{-} - ({\dot{X}}_{a})^{2}. 
} \eqno(19)
$$
and the conjugate momenta
$$\eqalign{
P_{a} &\equiv  {\dot{X}}_{a} {1 \over {\sqrt{\Delta}}} \sqrt{g}\cr
P_{+} &\equiv {1 \over {\sqrt{\Delta}}} \sqrt{g} .
} \eqno(20)
$$
Using these formulae for the conjugate momenta we can
evaluate the light-cone
Hamiltonian ${\cal{H}}_{M} \equiv  {\dot{X}}_{a} P_{a} + {\dot{X}}_{-} P_{+} - 
{\cal{L}}_{M}$ which equals 
$-\dot X_- {1 \over {\sqrt{\Delta}}} \sqrt{g}$ or  
$$
{\cal{H}}_{M} = {1 \over 2 P_{+}^\dagger}({|P_{a}|}^{2} + {1 \over 2} [X_{a}, 
X_{b}]^{2}).   
\eqno(21)
$$
We have defined  
$|P_{a}|^{2} \equiv P_{a}^{\dagger} P_{a}$
, ${1 \over P_{+}^\dagger} \equiv \sqrt{\Delta} {1 \over {\sqrt{g}}}$
 and used the fact that 
$$
( ({\dot{X}}_{a})^{2} + {\dot{X}}_{-} +
{1 \over 2} \Delta) {1 \over {\sqrt{\Delta}}} \sqrt{g} = 
{1 \over 2} \sqrt{\Delta} {1 \over {\sqrt{g}}} |P_{a}|^{2} .
$$ 
Note that the longitudinal matrix coordinate 
$X_{-}$ is eliminated via $2 {\dot{X}}_{-}
= -( \Delta + ({\dot{X}}_{a})^{2})$.
(Notice also that eqs. (18) - (21) resemble the analogous 
expressions in the continuum case, namely eqs. (4), (5), (7) and (8).)
 The expression (21) agrees with the bosonic
part of the Matrix theory Hamiltonian provided
that it is possible to take $P_{+}$ outside the trace (essentially
by stating that $P_{+} \propto {\bf 1}$), and by treating $P_{a}$ 
 as the conjugate momentum of $X_{a}$.
Furthermore,   
Matrix theory demands that the longitudinal momentum
$P_{+}$ be quantized $P_{+} = N/R$, $R$ being the extent
of the compact longitudinal direction.
How do we justify $P_{+} \propto {\bf 1}$? One
could argue that by rescaling the light-cone Hamiltonian
(21) by ${1 \over {\sqrt{\Delta}}} \sqrt{g}$, the longitudinal
momentum $P_{+} \propto {\bf 1}$, so that $P_{+}$ can 
be taken outside the
trace. One can then take the rescaled Hamiltonian to be the 
light-cone Hamiltonian, and define $P_{a}$ to be conjugate 
to $X_{a}$, thus making contact with Matrix theory in
the infinite momentum frame. 
Naturally, if the eigenvalues of
the longitudinal direction are all of order $R$, then 
$$
P_{+} \sim {1 \over R} {\bf 1}.  
 \eqno(22)
$$
 In other words, $Tr P_{+} \sim N/R$
(where $N$ defines the number of $D0-$branes as in Matrix theory).   

Let us recapitulate what we have done: By taking 
the existing covariant formulation of the
bosonic membrane we have attempted to formulate a covariant 
description of the bosonic part of Matrix theory. The crucial
issue was to come up with a well-motivated guess for the extension
of the original 
$U(N)$ symmetry that characterizes the 
infinite momentum frame limit of Matrix theory.
 Guided by the fact that
the infinite momentum frame membrane dynamics 
 is determined by the residual area preserving, or 
symplectic, diffeomorphisms, which form a subset of the original full 
three-dimensional
diffeomorphisms of the classical covariant membrane action, we
have enlarged the original $U(N)$ symmetry of Matrix theory 
by including 
"world-line" $t$-reparametrization invariance,  
thereby eliminating
the global time that is one of the defining features
of the 
infinite momentum frame formulation
of Matrix theory. It seems that it is necessary to let
$N \rightarrow \infty$ 
in order to define the matrix analog of the light-cone
gauge (15) and recover the global time. 

The crucial question arises: Can we really fix the gauge
in the large $N$ limit as indicated
by (15)? In other words, do we really have enough symmetry to
go to what we call matrix analog of the light-cone gauge?
These questions are crucially related to the problem of
a proper regularization of the original classical 
three-dimensional diffeomorphisms 
(also emphasized by 
M. Li and T. Yoneya in an unpublished work
[7]). 
 As recently pointed out by Smolin [6], the 
area preserving diffeomorphisms are realized linearly in
the classical membrane theory, and they
nicely map into $U(N)$ gauge transformations 
in Matrix theory (viewed as the 
quantization of the light-cone membrane theory). However, 
 the non-area preserving
diffeomorphisms, i.e. original three-dimensional
 diffeomorphisms modulo area
preserving diffeomorphisms, are realized non-linearly, and it is their 
regularization that is essential for the problem of regularization of
the full three-dimensional world-volume
 diffeomophisms. In the approach considered
in this article 
the non-area preserving diffeomorphisms are not 
explicitly taken into account.

Therefore it seems that in order to answer the crucial questions pertaining 
to the problem of a covariant formulation of the bosonic 
 Matrix
theory, to wit: 
the enlargement of the original $U(N)$ gauge symmetry (or put differently,
the proper "quantization" of the three-dimensional diffeomorphisms of the
covariant formulation of the bosonic membrane), the issue of gauge
fixing and recovery of the $U(N)$ invariance in the infinite momentum
limit, the decoupling of the longitudinal matrix coordinate $X_{-}$,
 and the
quantization of the longitudinal momentum $P_{+}$, one
should first understand the question of "quantization" of the non-area
preserving diffeomorphisms in the classical membrane 
theory.
It is not clear at the moment how any of the above issues
are affected by supersymmetry, which is surely the most important aspect 
of the physics of Matrix theory in its current form.
(We add that we do not know whether the question of locality can be
addressed in the present approach, nor whether can one construct many-body
states from block diagonal matrices as in [2]. Also it is not
clear what kind of objects replace partons in the covariant approach,
and what role, if any, is played by the holographic principle [2], [8].) 

Finally, let us briefly note that 
the real covariant formulation of Matrix theory
should naturally incorporate the space-time uncertainty
principle of Yoneya [9], and Li and Yoneya [10], 
which is known to hold both in 
string theory and Matrix theory.
Within the approach outlined above the space-time uncertainty
principle should be expected to come out from the following
commutator 
$$
[X_{+}, X_{a}] \sim  l^2, \eqno(23)
$$
where $l$ denotes the fundamental unit of length.
 The expression (23) naturally
 leads to 
$$
\delta t \delta x \sim l^2, \eqno(24)
$$
which represents the usual formulation
of the space-time uncertainty principle [9],[10].

\vskip .1in
{\bf Acknowledgements}

We are grateful to T. Banks, P. Berglund, S. Chaudhuri,
J. Harvey, M. Gunaydin, A. Klemm, R.Leigh, J. Polchinski,
M. Rozali, 
P. Pouliot, L. Smolin and especially M. Li and T. Yoneya for many important
discussions. D. M. wishes to thank the 
 Institute for Theoretical Physics, Santa Barbara and
Enrico Fermi Institute of the  University of Chicago, 
for providing such stimulating working environments 
during the time when most of
this work was done. We are deeply indebted to
 S. Chaudhuri for many useful suggestions
regarding the original version of this note. 
This work is supported in part by 
Grant-in-Aid for Scientific Research from Ministry of Science and Culture.

\vskip.1in
{\bf References}

\item{1.} For references consult 
B. de Wit, J. Hoppe and H. Nicolai, Nucl. Phys. B305 (1988) 545.
\item{2.} T. Banks, W. Fischler, S. H. Shenker and L. Susskind, Phys. 
Rev. D55
(1997) 5112. For a review and further references,
 see T.Banks, "Matrix Theory", hep-th/9710231.  
\item{3.} J. Goldstone, unpublished; J. Hoppe, MIT Ph.D.
 thesis, 1982 and in
 "Proc. Int. Workshop on Constraint's Theory and
 Relativistic Dynamics", G. Longhi and L. Lusanna,
 eds. (World Scientific, 1987); J. Hoppe, Int. J.
 Mod. Phys. A4 (1989) 5235;
 D. Fairlie, P. Fletcher and C. Zachos, J. Math. Phys. 31 (1990) 1088.
\item{4} J. Polchinski, Phys. Rev. Lett. 75 (1995) 4724;
J. Polchinski, S. Chaudhuri and C. V. Johnson,
"Notes on $D$-branes, hep-th/9602052; J. Polchinski, 
"TASI Lectures on $D$-branes", hep-th/9611050; E. Witten, Nucl. Phys.
B460 (1995) 338; M. Douglas, D. Kabat, P. Pouliot, S, Shenker,
Nucl. Phys. B485 (1997) 85.   
\item{5.} J. B. Barbour, Nature 249 (1974) 328 (Erratum, Nature 250
(1974) 606); Nuovo Cimento 26B (1975) 16; J. B. Barbour and B. Bertotti,
Nuovo Cimento 38B (1977) 1; Proc. Roy. Soc. Lond. A 382 (1982) 295.
\item{6.} L. Smolin, "Covariant Quantization of Membrane Dynamics", 
hep-th/9710191.
\item{7.} M. Li and T. Yoneya, private communication.
\item{8.} L. Susskind, J. Math. Phys. 36 (1995) 6377; G. 't Hooft,
"Dimensional Reduction in Quantum Gravity". gr-qc/9310026.  
\item{9.} T. Yoneya, Mod. Phys. Lett. A4 (1989) 1587. See also
T. Yoneya, in "Wandering in the Fields". K. Kawarabajashi and A. Ukawa, 
eds.
(World Scientific, 1987) pp.419; and "Quantum String Theory", N. Kawamoto 
and
T. Kugo, eds. (Springer, 1988) pp.23.
\item{10.} M. Li and T. Yoneya, Phys. Rev. Lett. 78 (1997) 1219.

\end